# RAIDER: Rapid, anatomy-independent, deep learning-based PDFF and $R_2^*$ estimation using magnitude-only signals, dual neural networks and training data distribution design

Timothy JP **Bray** [1,2,3], Giulio V **Minore** [3], Alan **Bainbridge** [4], Louis **Dwyer-Hemmings** [1,2,3], Stuart A **Taylor** [1,2], Margaret A **Hall-Craggs** [1,2], Hui **Zhang** [3]

**1** Centre for Medical Imaging, University College London, UK
**2** Department of Imaging, University College London Hospital, UK
**3** The Hawkes Institute, University College London, UK
**4** Department of Medical Physics, University College London Hospital, UK

**Abstract**
There has been recent interest in the use of magnitude-based fitting methods for estimating proton density fat fraction (PDFF) and $R_2^*$ from chemical shift-encoded MRI (CSE-MRI) data, since these methods can still be used when complex-based methods fail or when phase data are inaccessible or unreliable (such as in multicentre trials, low resource settings, preclinical imaging and national cohort datasets), and may also be used as a final processing step with complex-based methods. However, conventional fitting techniques are computationally expensive. Deep learning (DL)-based methods promise to accelerate parameter estimation, but there are no existing deep learning methods for voxelwise CSE-MRI tissue parameter estimation. Here, we show that a naive voxelwise multilayer perceptron (MLP) implementation suffers from poor performance because multiple tissue parameter values can produce similar signals (degeneracy), potentially accounting for this gap in the literature. To address this problem and realise the potential acceleration offered by MLP-based tissue parameter estimation, we propose RAIDER, a voxelwise method for **r**apid, **a**natomy-**i**ndependent **de**ep lea**r**ning-based PDFF and $R_2^*$ estimation using multi-echo magnitude-data. RAIDER utilises two neural networks, each with a separately-designed training data distribution, to deal with degeneracy and thus realise the benefits of an MLP-based approach, with a 400-fold to 2800-fold acceleration (meaning that processing takes just seconds, rather than minutes or hours). Our code is available at `https://github.com/TJPBray/DixonDL`.

Corresponding author: t.bray@ucl.ac.uk

## 1. Introduction

**Chemical** shift-encoded MRI (CSE-MRI) is a reliable, fast method for quantifying proton density fat fraction (PDFF) in a wide variety of organs and disease states (Reeder et al., 2004, 2005; Hernando et al., 2008; Liu et al., 2007; Noureddin et al., 2013; Bray et al., 2018b). PDFF measurements are now established for the measurement of steatosis in liver (Noureddin et al., 2013), and are also increasingly used for other organs including pancreas (Yoon et al., 2016; Kuhn et al., 2015), muscle (Morrow et al., 2016) and bone marrow (Bray et al., 2018a; Latifoltojar et al., 2017; Messiou et al., 2019). CSE-MRI most commonly utilises multi-echo gradient echo acquisitions, meaning that $R_2^*$ measurements can be extracted from the same acquisition, providing additional quantitative information on iron or calcium. PDFF and $R_2^*$ measurements estimated using gradient echo acquisitions therefore offer flexible biomarkers informing on a wide variety of pathological processes. The speed of the underpinning acquisitions means that these biomarkers can also be acquired in the context of whole body MRI (WB-MRI), which is becoming an important tool for lesion characterisation and disease staging in cancer (Messiou et al., 2019; Taylor et al., 2019; Costelloe et al., 2012), amongst other applications such as the assessment of inflammatory arthritis (Choida et al., 2022, 2024).

From a signal processing perspective, the central challenge of CSE-MRI is to disentangle or separate the signals arising from water and fat. This is challenging because the signal arising from water-dominant and fat-dominant tissues is very similar, giving rise to the so-called 'fat-water ambiguity' problem. The prevailing approach to solving this problem has been to use the phase of the complex signal in order to separate water and fat: with complex signal-based methods, water-dominant and fat-dominant tissues can be distinguished on the basis of their phase, so long as the contribution to the phase from main magnetic field $B_0$ can be estimated accurately and therefore separated from the phase arising due to chemical shift (Reeder et al., 2004, 2005; Hernando et al., 2008; Liu et al., 2007; Noureddin et al., 2013).

However, although these complex signal-based methods can be very effective, in some situations complex data (and/or expensive vendor-supplied software packages for PDFF estimation) may be unreliable or not available, creating a substantial barrier to the use of CSE-MRI. This problem applies to various scenarios including multi-centre studies (Bainbridge et al., 2020; Eck et al., 2024; Wong et al., 2023; Waddell et al., 2022; Losev et al., 2024), low resource clinical and research settings (Murali et al., 2022), preclinical imaging (Ghimire et al., 2025) and when retrospectively analysing existing datasets such as the UK biobank and German national cohort study (Littlejohns et al., 2020; Consortium, 2014).

As a result of these limitations, a different class of methods using only the magnitude of the signal (i.e. where the signal phase is discarded) for tissue parameter estimation have been developed. As phase information is not available to these methods, they require a different source of information to resolve fat-water ambiguity. Recently, the development of the MAGnitude Only (MAGO) and MAGnitude Only with RIcian NOise modelling (MAGORINO) methods demonstrated that this ambiguity could be resolved on the basis of subtle differences in the oscillations of the magnitude signal over multiple echoes (Triay Bagur et al., 2018; Bray et al., 2023). In addition to being usable in situations where complex data and/or expensive vendor-supplied software packages are not available, these methods also avoid phase errors which can induce bias when fitting with complex data, meaning that they may be more accurate and robust than complex signal-based techniques in a real-world setting. Related to these advantages, the MAGO technique (not developed by these authors) has achieved regulatory clearance as a commercial software tool for PDFF and $R_2^*$ quantification in the liver within the LiverMultiscan package (Perspectum, Oxford) and is increasingly being used in clinical trials (Wong et al., 2023; Waddell et al., 2022; Losev et al., 2024). Importantly, magnitude-based fitting is also commonly used as a final processing step even in complex fitting pipelines (Hernando et al., 2017), and is the central fitting method used in the Siemens qDixon package (Zhong et al., 2014). There has also been interest in using magnitude-fitting as a first processing step in order to improve estimation of $B_0$ inhomogeneity, with a view to performing quantitative susceptibility mapping (QSM) (Triay Bagur et al., 2023).

These developments suggest that magnitude-based fitting will continue to have a crucial role in CSE-MRI processing. However, these methods carry a substantial computational cost, due to the iterative nature of the algorithms as well as the need to perform fitting with nonlinear least squares several times - with different start points - in order to ensure that the correct likelihood maximum / error minimum is found (Triay Bagur et al., 2018; Bray et al., 2023). This increases the time, human resource and funding required to use these techniques, limiting their use in research settings and in clinical practice.

A promising approach to reducing this computational cost is to use deep learning for tissue parameter estimation: this has the advantage that computational cost is front-loaded: training can be performed in advance, before images are acquired, and the computational resources needed to actually process those images are typically much less than with conventional fitting (Barbieri et al., 2019; Epstein et al., 2024; Gyori et al., 2021) (note that these methods use MLPs and take the signal from an individual voxels as input, in contrast to convolutional neural network (CNN) -based methods that take images or parts of images as input - the differences between these approaches are considered in more depth in the Discussion section). However, to our knowledge there is no existing voxel-independent deep learning-based method for PDFF and $R_2^*$ fitting of multi-echo CSE-MRI data.

To address this problem, here we develop a novel approach using multiple multilayer perceptrons (MLPs) for PDFF and $R_2^*$ estimation from multi-echo magnitude signals. We first identify a crucial problem with a 'naive' single MLP implementation, related to model degeneracy: the network performs poorly because multiple combinations of tissue parameter values can produce very similar signals. To solve this degeneracy problem, we introduce a novel method whereby two MLPs, each trained separately with a restricted training distribution, are used to avoid degeneracy and thus estimate these tissue parameters. The new method is referred to as RAIDER: **r**apid, **a**natomy-**i**ndependent **de**ep lea**r**ning-based CSE-MRI.

## 2. Theory and related works

### 2.1 Tissue and noise models

A full description of the tissue and noise models used in this work is given in Bray et al. (2023); the key tissue and noise models are briefly summarised here.

With a gradient echo-based CSE-MRI acquisition, the noise-free complex signal $S$ acquired at echo time $t$ can be modelled as:

$$S(t|S_0, PDFF, R_2^*, f_B) = S_0\left((1-PDFF) + PDFF\sum_{q=1}^{Q} r_q e^{i2\pi f_{F,q}t}\right) e^{i2\pi f_B t} e^{-tR_2^*} \quad (1)$$

where $S_0$ and $PDFF$ are the theoretical signal intensity at t=0 and proton density fat fraction respectively, $f_{F,q}$ is the frequency of each spectral fat component, $r_q$ is the relative amplitude of each spectral fat component, $Q$ is the total number of spectral fat components, $f_B$ is the frequency offset due to $B_0$ inhomogeneity and $R_2^* = 1/T_2^*$ is an unknown relaxation constant. It is conventional to assume that the relative amplitudes and frequency offsets of each fat component are known *a priori*; therefore the unknown tissue parameters are $S_0$, $PDFF$, $f_B$ and $R_2^*$. Note that this formulation slightly differs from that in Bray et al. (2023), where the tissue parameters included $\rho_f$ and $\rho_w$ rather than $S_0$ and $PDFF$. The current formulation has been chosen as it allows a more convenient estimation of $S_0$ after the DL-based estimation of PDFF and $R_2^*$, as detailed in Section 3.1.4.

The frequency offsets and relative amplitudes for the multipeak fat spectrum were matched to Triay Bagur et al. (2018); Bray et al. (2023); Hernando et al. (2008), i.e. we used frequency shifts relative to the water peaks of 3.90, 3.50, 2.70, 2.04, 0.49, and +0.50 ppm and relative amplitudes of 0.087, 0.694, 0.128, 0.004, 0.039, and 0.048).

In the presence of Gaussian noise (present in both real and imaginary channels), the log likelihood for a set of predicted noise-free signals, given the measured noisy signals, is given by

$$\log L(\{S_i\}, \sigma^2|\{S_i'\}) = -n\log(\sqrt{2\pi\sigma^2}) - \sum_{i=1}^{n}\frac{|S_i' - S_i|^2}{2\sigma^2} \quad (2)$$

where $\{S_i\}$ is the set of predicted signals based on the tissue parameter estimates, $\{S_i'\}$ is the set of measured signals, $\sigma^2$ is the variance of the Gaussian distribution for the noise, $n$ is the number of measurements (double the number of echo times when using the complex signal).

For the signal magnitude, the noise-free signal in Eqn (1) becomes

$$M(t|S_0, PDFF, R_2^*) = S_0\left|(1-PDFF) + PDFF\sum_{q=1}^{Q} r_q e^{i2\pi f_{F,q}t}\right| e^{-tR_2^*} \quad (3)$$

with only three unknown tissue parameters: $S_0$, $PDFF$, and $R_2^*$.

Taking the magnitude dictates that each noisy measurement now has a Rician distribution. The log likelihood for a set of predicted noise-free signals becomes

$$\log L(\{M_i\}, \sigma^2|\{M_i'\}) = \sum_{i=1}^{n}\left[\log\frac{M_i'}{\sigma^2} - \frac{M_i'^2 + M_i^2}{2\sigma^2} + \log I_0(\frac{M_i' M_i}{\sigma^2})\right] \quad (4)$$

where $\{M_i'\}$ is the set of measured magnitude signals at different echo times, $\{M_i\}$ is the corresponding set of predicted magnitude signals and $I_0$ is the $0^{th}$ order modified Bessel function of the first kind.

### 2.2 Dual optima problem and the source of degeneracy in CSE-MRI

With magnitude-based fitting, the likelihood functions defined in Eqns (2) and (4) have two optima. In the absence of noise, there is one global optimum that corresponds closely to the ground truth tissue parameters ('true solution') and one local optimum with a PDFF value at the opposite end of the range (i.e. with a PDFF value close to $1 - PDFF_{true}$), with incorrect or 'swapped' tissue parameters ('swapped solution') (Fig. 1). When the signal becomes noisy, the global optimum will correspond to the true solution more often than not, but in some voxels the global optimum will correspond to the incorrect/swapped solution (the frequency with which this occurs depends on the signal-to-noise ratio). This creates a 'one-to-many' problem whereby the same signal can be produced by two different parameter combinations. This is a degeneracy problem of the sort highlighted in Section 2.4 and is one which standard deep learning methods are poorly equipped to solve.

Note that conventional fitting methods such as MAGO and MAGORINO (Triay Bagur et al., 2018; Bray et al., 2023) address the issue of having multiple optima in the likelihood function by using two-point search to ensure that both optima are explored, and comparing the likelihood of the two optima in order to find the correct tissue properties. With these methods, there is a stationary or 'switching' point close to 60% in the PDFF dimension, below which the initial guesses are likely to converge to the low PDFF optimum and above which the initial guesses are likely to converge to the high PDFF optimum (in 2D this is a saddle or minimax point) - this switching point is shown in Figure 1. This issue has also been discussed in prior work (Triay Bagur et al., 2018; Bray et al., 2023) and informs the partition of the training distribution described in Figure 2.

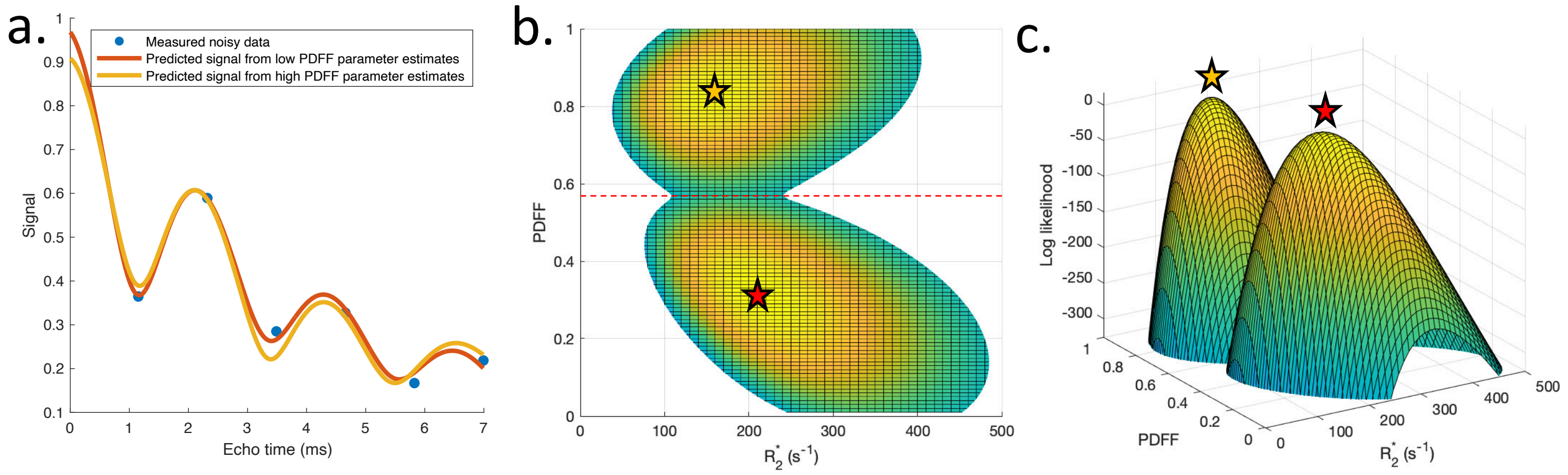

Figure 1: Illustration of how the presence of two likelihood optima leads to degeneracy in CSE-MRI. The measured signal (a) and Rician log likelihood (b, top-down view, c, oblique view) as a function of PDFF and $R_2^*$ are shown for a typical noisy simulated voxel with PDFF $= 0.3$ and $R_2^* = 200$ s$^{-1}$. In this voxel, there are two sets of parameter estimates producing very similar signal (a), and with very similar likelihood (b,c). One of these parameter sets [red line (a), red star (b,c)] is the global optimum, corresponding to the true low-PDFF parameter estimates, and one [yellow line (a), yellow star (b,c)] is a local optimum at high PDFF, corresponding to an incorrect or 'swapped' high PDFF set of parameter estimates. However, in a minority of other voxels the the global optimum can correspond to the incorrect, swapped parameter estimates. This creates a source of degeneracy - for a given set of signals, there are two 'candidate solutions' which have very similar likelihood, creating a 'one-to-many' problem which traditionally deep learning approaches are poorly equipped to solve. The switching line at PDFF $= 0.58$ (explained in Section 3.2.2) is shown as a dotted red line. In (b) and (c), likelihood values below the lower limit of the y-axis are not shown to improve the clarity of the figure.

## 2.3 Tissue parameter estimation with deep learning

The use of neural networks for curve fitting was first proposed by Bishop and Roach (1992), and has become increasingly widely used for qMRI tissue parameter estimation with the arrival of deep learning. Unlike with conventional fitting, which iteratively finds the points on an objective function at which error is minimized / likelihood is maximised, a deep neural network is trained to directly map a single voxel's signal to its corresponding qMRI tissue parameters (Golkov et al., 2016). The unknowns in this model are thus the network weights rather than the parameters in the tissue model; the training procedure aims to find the values of these weights that minimise the distance between the predicted tissue parameter estimates and the training labels (ground truth tissue parameter values) in the training dataset. Once the network has been trained, tissue parameter estimation for any data can be achieved in a single shot by inputting the data to the network. This effectively front-loads the computational cost and means that the cost at the time the network is applied is dramatically reduced compared to conventional fitting.

## 2.4 Degeneracy in deep learning

Bishop noted that neural networks were able to represent one-to-one and many-to-one mappings, but could not represent one-to-many ('multivalued') mappings (Bishop and Roach, 1992). This problem was referred to by Bishop as an 'ambiguity phenomenon' and is referred to here as degeneracy. In Bishop's experiments, the quality of the neural network fits was found to be very poor in regions of tissue parameter space where this degeneracy problem occurred. Guerreri et al. recently provided further theoretical insights into this problem, showing that, in the presence of degenerate samples during supervised training, the function learned by the neural network will map the degenerate signal to the empirical mean of tissue properties over the degenerate subset (Guerreri et al., 2023). In other words, in this situation the network does not learn either of the candidate solutions.

A problem of this type is encountered in CSE-MRI, where a one-to-many problem exists because there are two sets of tissue parameter values (the 'true' solution and the 'swapped' solution) which correspond to very similar signals. This problem is relatively minor at low $R_2^*$ but becomes more severe at higher $R_2^*$ and/or lower SNR, as the water and fat peaks become broader and therefore more difficult to resolve (Bray et al., 2023). Bishop suggested that, in cases where there were multiple potential solutions for a given input, this problem could be solved by excluding a region of the output space (Bishop and Roach, 1992). However, this approach assumes that one region of output space can be discarded on the basis that it is unlikely to occur in experiments. With CSE-MRI, this is not a satisfactory approach because both low PDFF and high PDFF values can be present in real tissue.

To address this problem, we propose dividing the tissue parameter space into two parts, in such a way that one network can be trained on the first part of tissue parameter space, and another network can be trained on the second part of tissue parameter space, without degeneracy. Specifically, we divide the tissue parameter space into low PDFF regions, with PDFF values below the switching line, and high PDFF regions, with PDFF values above the switching line. The value of the switching line is chosen by inspecting the objective functions shown in Figure 1 to determine the location of the switching point in the PDFF dimension, as described previously (Triay Bagur et al., 2018; Bray et al., 2023); further implementation details are provided in Section 3.2.2.

By dividing the training distribution in this way and training separate networks, we create two networks that are each accurate within the region of tissue parameter space on which they are trained. The correct output from the two networks can then be chosen based on the difference in likelihood (as defined in Eqn (4)) between the two solutions. Our approach is analogous to conventional fitting approaches where different optima ('candidate solutions') are explored using multiple start points to ensure that the global optimum (and the correct solution) is found (Triay Bagur et al., 2018; Bray et al., 2023), but carries an almost negligible computational cost.

## 3. Methods

### 3.1 Network

An overview of the training procedures for single and dual network approaches and the inference procedure for the dual network approach is provided schematically in Figure 2; detail on, and justification of, the individual elements within this schematic is provided in the following subsections.

#### *3.1.1 Normalization using single peak approximation of $S_0$*

Unlike existing voxelwise DL methods focusing on diffusion MRI (Karimi, 2024), where b=0 is directly measured, with CSE-MRI we do not have access to a signal measurement at TE=0, meaning that a simple normalization method is not readily available. To address this problem, and to make network predictions robust to variations in signal intensity (i.e. variations in $S_0$), we developed a novel normalization procedure: the multi-echo signals were normalized by a rough approximation of $S_0$ before being inputted to the network(s). This rough $S_0$ approximation, denoted $\hat{S}_0$, was obtained by taking the measured signal for the two echo times closest to the first and second 'in-phase' echo times (assuming a single peak tissue model), calculating the decay between these echo times (assuming monoexponential decay) and then extrapolating back to $t = 0$ using a log-linear fit. Note that these chosen echo times do not vary between voxels but are protocol-specific. To ensure that this procedure did not result in an $\hat{S}_0$ lower than the maximum measured signal intensity, if $\hat{S}_0 < \max M(t)$, then $\hat{S}_0$ was set to be equal to be $\max M(t)$.

This step was applied both during training and at inference (see Figure 2). Note that this step is not expected to produce accurate tissue parameter estimates, and it is anticipated that the network learns to compensate for inaccuracies in this step. Note also that a further, updated estimate of $S_0$ is obtained after the network predictions, as detailed in 3.1.4.

#### *3.1.2 Network architecture*

We used a simple network design consisting of five fully-connected layers, with first four having the same number of nodes as the number of signal measurements (i.e. echo times), and the fifth (output) layer having a node for each tissue parameter (i.e. one output node for PDFF and one for $R_2^*$), with exponential linear unit (ELU) activation functions (Clevert et al., 2015; Epstein et al., 2024). This design is similar to Barbieri et al. (2019) and Epstein et al. (2024), but uses five layers rather than three to increase network capacity.

To ensure that these networks had sufficient capacity, we also implemented several deeper and wider networks, the largest of which had fifty times as many nodes as the number of signal measurements in each fully connected hidden layer and twelve rather than five fully connected layers (11 hidden layers and one output layer). The capacity of both standard and larger networks was examined in experiments analogous to those described in 3.2 but using noise-free training data, as detailed in Appendix A. These results (taken together with the rest of the experiments) show that larger networks offer a performance benefit, due to greater capacity, in the noise-free setting but that the smaller networks show equivalent performance when training and evaluation are performed in the presence of noise. The finding that greater capacity is needed for noise-free data is consistent with Gyori et al. (2021). The networks were implemented in Matlab 2021b using the MATLAB Deep Learning toolbox.

#### *3.1.3 Network training*

The network was trained in a supervised fashion using entirely simulated data based on the tissue models highlighted in Section 2.1. First, tissue parameter values (labels) were randomly drawn from PDFF / $R_2^*$ tissue parameter space (using a uniform distribution within the defined tissue parameter ranges to minimise bias (Gyori et al., 2021)); the corresponding simulated signal intensities were then simulated, and Rician noise was added with SNR values randomly sampled from a uniform distribution (Mastropietro et al., 2019) between 20 (very low SNR) and 120 (very high SNR, designed to exceed the upper end of the SNR range expected in vivo (Triay Bagur et al., 2018; Bray et al., 2023)). Normalisation was performed as detailed in 3.1.1 and as shown in Figure 2.

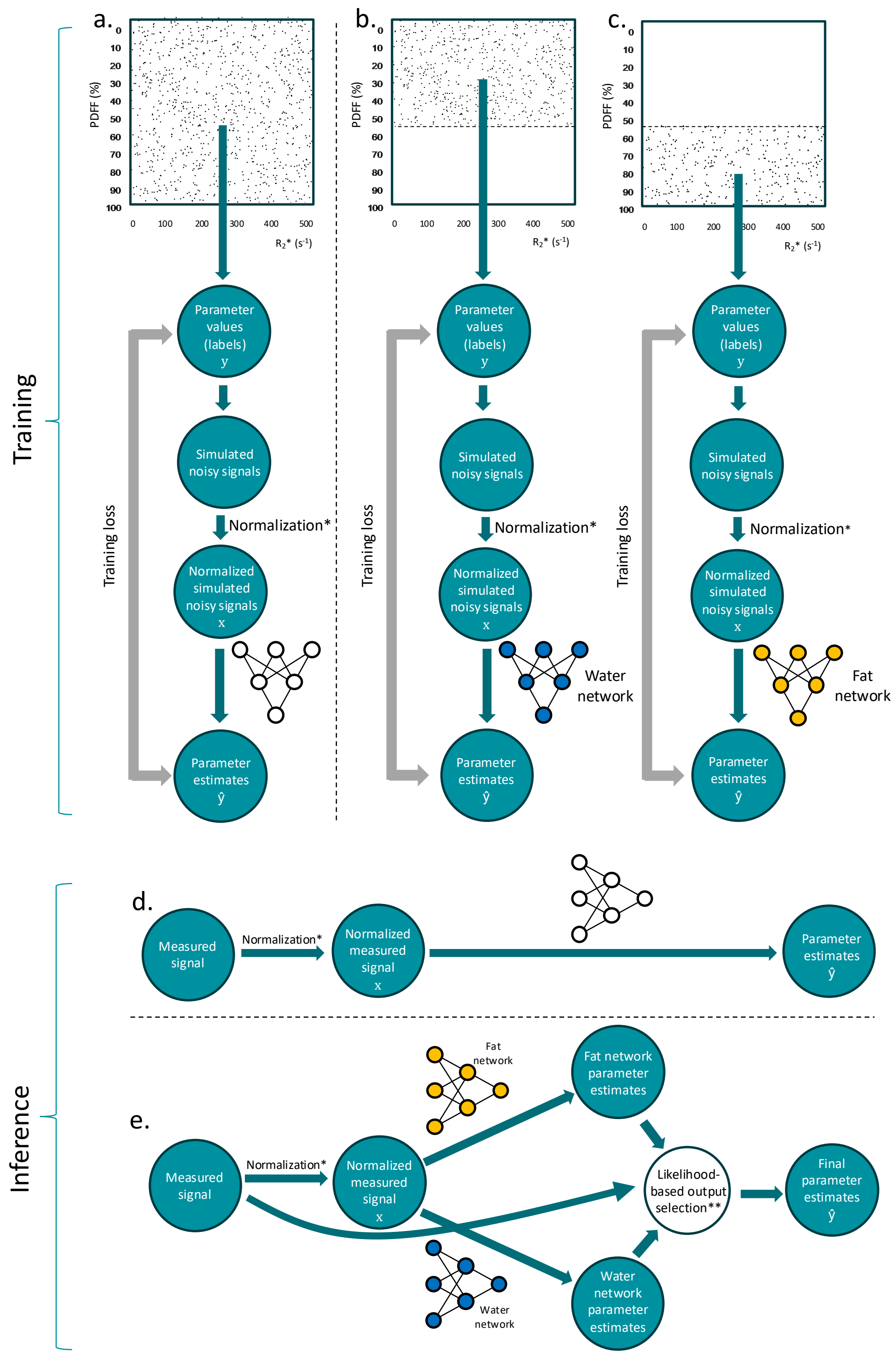

Figure 2: Training procedure for single (a) and dual (b,c) network approaches, and inference procedure for single network (d) and dual network (e) approaches. With a naive single network approach (a), one network is trained with a uniform distribution across the entire space of plausible tissue parameter values. With the proposed approach, two separate networks are trained on separate parts of the plausible tissue parameter space to avoid degeneracy: a 'water network' (b) is trained only on the region of tissue parameter space below the 'switching line' (dashed line), and a 'fat network' (c) is trained only on the region of tissue parameter space above the switching line (dashed line). At inference, for the dual network approach (e), the outputs from the two networks (fat network and water network) are treated as 'candidate' solutions; the likelihood for each candidate is calculated based on the measured signal and the one with the higher likelihood is selected (e). *Normalization is performed as detailed in 3.1.1. **The selection of either the water network or fat network output is performed as described in 3.1.4.

Training was performed using an Adam optimiser with learning rate 0.001 (Epstein et al., 2024), minibatch size 32, validation patience 50 (more conservative than Barbieri et al. (2019), who used validation patience of 10). Additionally, to enable effective training of the larger networks, for these networks we used a piecewise learning rate schedule with drops in the learning rate by a factor of 0.2 every 5 epochs, and did not use validation patience. L2 regularization was not used as the size of the synthetic dataset was deemed sufficient to minimise overfitting. Training was performed using a training dataset of 100,000 samples (Epstein et al., 2024); in the single network each network was trained on the full training dataset, whereas in dual network case the dataset was first subdivided, as detailed in 3.2.2. In both cases, the datasets were split 80/20 between training and validation. Training was performed on an Apple iMac with 3.8-GHz, using a single central processing unit (CPU). Further details of network training are detailed in the individual experiments below. Network training took approximately 40 minutes in total (20 minutes per network) for the proposed two network approach shown in Figure 2.

#### *3.1.4 Network output selection based on likelihood*

To determine which of the two networks' predictions should be taken as the final output, the likelihood for the predicted signals from each network, given the measured signals, was compared.

To enable this, we first reformulate Eqn (3) as

$$M(t|S_0, PDFF, R_2^*) = S_0 a_t \tag{5}$$

where $a_t$ is a known real-valued scalar quantity, calculated from the RAIDER tissue parameter estimates using

$$a_t = |(1 - PDFF) + PDFF \sum_{q=1}^{Q} r_q e^{i2\pi f_{F,q} t}| e^{-tR_2^*}. \tag{6}$$

Once $a_t$ has been calculated, a new, improved estimate of $S_0$, $\tilde{\hat{S}}_0$, (note that this is distinct from the earlier, rough approximation $\hat{S}_0$) at the maximum Gaussian likelihood can be obtained using the matrix formulation

$$\mathbf{m} = \mathbf{a}\tilde{\hat{S}}_0 \tag{7}$$

where $\mathbf{m}$ is an N x 1 vector of measured signal magnitudes and $\mathbf{a}$ is an N x 1 vector containing the values of $a_t$ for each sampled echotime, and N is the total number of sampled echo times. N can be the total number of echo times acquired (we used this approach here), or a smaller number of echo times to reduce the impact of Rician noise on the estimates (we did not find any improvement in performance for this approach).

$\tilde{\hat{S}}_0$ can be estimated using

$$\tilde{\hat{S}}_0 = (\mathbf{a}^\top \mathbf{a})^{-1} \mathbf{a}^\top \mathbf{m} \tag{8}$$

Having calculated $S_0$, the log likelihood for the predicted signals, based on the measured signals, was calculated using Eqn (4) for each of the two networks' tissue parameter estimates, and the network predictions giving the higher likelihood were taken as the output.

To facilitate selection of the correct network output in the occasional situation that one network produced high likelihood but physically implausible tissue parameter estimates (specifically negative $R_2^*$ values, which could occur when a network was exposed to signals outside its training distribution), such negative $R_2^*$ values were deemed to be incorrect, and the outputs of the other network were selected.

### 3.2 Simulation Experiments

A series of simulation experiments were conducted to investigate the effect of training data distribution on the accuracy and precision of tissue parameter estimation.

The training data was constructed by simulating noisy multi-echo signals using Eqn (3) for a range of PDFF and $R_2^*$ values, using a typical CSE-MRI acquisition protocol with six echoes: $TE_1 = 1.152$ ms and $\Delta TE = 1.168$ ms, matching the first 3T protocol in Hernando et al. (Hernando et al., 2017). The distribution of PDFF and $R_2^*$ values was varied for each experiment by defining regions of tissue parameter space from which the PDFF and $R_2^*$ values could be selected, as detailed below and in Figure 2. PDFF and $R_2^*$ values were randomly selected within the defined region.

Performance was evaluated on a separate simulated test dataset consisting of evenly-spaced PDFF and $R_2^*$ values across the full tissue parameter space (PDFF $\in [0, 1]$, $R_2^* \in [0, 500]$ s$^{-1}$ ), resulting in 101 x 21 = 2121 combinations of tissue parameter values. For each combination of tissue parameter values, we simulated 100 noisy signals (at SNR = 60, which is typical for CSE-MRI) in order to mimic repeating the same acquisition 100 times. Performance was evaluated in terms of the bias and precision of tissue parameter estimates, and compared against magnitude fitting with the MAGORINO algorithm (Bray et al., 2023).

#### 3.2.1 Single network with uniform training distribution

The training dataset consisted of 100,000 sets of magnitude signals derived from randomly-generated tissue parameter combinations with a uniform distribution in PDFF / $R_2^*$ tissue parameter space (PDFF $\in$ $[0, 1]$, $R_2^* \in$ $[0, 500]$ $s^{-1}$ ), as demonstrated in Figure 2(a).

#### 3.2.2 Dual networks with modified training distribution

To address the effect of degeneracy on network predictions, the full dataset consisting of 100,000 signal sets was divided into two separate, smaller training datasets, as demonstrated in Figure 2(b,c):

The first network ('water network') was trained only on the low PDFF training data, i.e. with PDFF values below the switching line (PDFF $\in$ $[0, 0.58]$, $R_2^* \in$ $[0, 500]$ $s^{-1}$ ). Details on determining the value of the switching line are detailed in 2.4.

The second network ('fat network') was trained only on the high PDFF training data, i.e. with PDFF values above the switching line (PDFF $\in$ $[0.58, 1]$, $R_2^* \in$ $[0, 500]$ $s^{-1}$ ).

This approach - dividing the training data seen by the single network into two smaller datasets - (i) ensures that the dual network approach has seen the same training data as the single network approach and (ii) ensures that the tissue parameter space sampling density is kept the same for all networks.

Having trained these two networks and obtained the predictions on the test dataset, a further set of 'composite' predictions was derived by choosing the output from the two networks with the highest likelihood, as described in Section 3.1.4.

### 3.3 Phantom experiments

To evaluate the feasibility of DL-based fitting across a range of scanners and protocols, our method was evaluated in a publicly available multisite, multivendor, and multi-field-strength phantom data set - full details of the dataset are given in Hernando et al. (2017). Briefly, the dataset consists of fat–water mixtures with known varying fat fraction, scanned at 1.5T and 3T at six centers (two centers each for GE, Philips, and Siemens). Data acquisition was performed using each site's version of a multi-echo 3D spoiled gradient echo CSE sequence (six echo times), with two different protocols (each performed at 1.5T and 3T) performed in order to test the reproducibility across different acquisition settings. Protocols 1 and 2 were performed at 1.5T, and protocols 3 and 4 were performed at 3T. Protocols 1 and 3 generated approximately in-phase and opposed-phased echoes, whereas protocols 2 and 4 used the shortest echoes achievable.

Deep learning-based tissue parameter estimation was performed as described in Section 3.2.2; new network models were trained for each of the acquisition protocols used (again, the training was performed using only simulated data, meaning that all *in vivo* data was entirely unseen at inference). After generating tissue parameter maps, fat fraction values were obtained from the individual vials of the phantom by taking the median value from ROIs placed on the individual tubes (Triay Bagur et al., 2018; Bray et al., 2023) and compared against known reference values. The performance of the networks was compared against conventional fitting.

### 3.4 In vivo imaging

To evaluate the feasibility of the proposed approach *in vivo*, and to demonstrate anatomical robustness, we imaged multiple different anatomical regions in two separate subjects, on different scanners. The first subject had their pelvis (with the acquisition focusing on the sacroiliac joints) imaged on a 3T Philips Ingenia system using a multi-echo 3D spoiled gradient echo sequence with flyback gradients and monopolar readouts $TE_1 = 1.2$ ms, $\Delta TE = 1.6$ ms, six echoes, flip angle = 5°, TR = 25 ms, matrix size = 320 × 320, and pixel spacing = 1.8 × 1.8 mm. The second subject had their thorax, pelvis and abdomen imaged on a Siemens 3T Vida scanner using a similar acquisition with $TE_1 = 1.1$ ms, $\Delta TE = 1.1$ ms. Deep learning-based tissue parameter estimation was performed as described in Section 3.2.2; again, for each of the acquisition protocols used, new network models were trained using purely simulated data, meaning that all *in vivo* data was entirely unseen at inference. Conventional magnitude-based fitting was performed using both Gaussian and Rician noise models, as described previously (Triay Bagur et al., 2018; Bray et al., 2023). These scans were performed with institutional review board approval (Queen Square Research Ethics Committee, London, REC 15/LO/1475), and subjects provided written informed consent.

## 4. Results

### 4.1 Simulation experiments

#### 4.1.1 Single network with uniform training distribution

Figure 3 (left column) shows the PDFF estimation performance of a single network trained with a uniform distribution of tissue parameter values over the full tissue parameter space, evaluated in the presence of noise. At low PDFF, there are regions of positive error (shown in yellow in (f)) towards the right of subplots (a) and (f) (i.e. with increasing $R_2^*$), with estimates becoming closer to the mean PDFF. Conversely, at high PDFF, there are regions of negative error (shown in blue in f) with increasing $R_2^*$, again with estimates becoming closer to the mean PDFF. The precision of tissue parameter estimates (k) also worsens with increasing $R_2^*$. The second column shows very little change in the results for a substantially larger network (with approximately 2500 times as many learnable parameters), indicating that the poor performance of this network is not due to lack of network capacity.

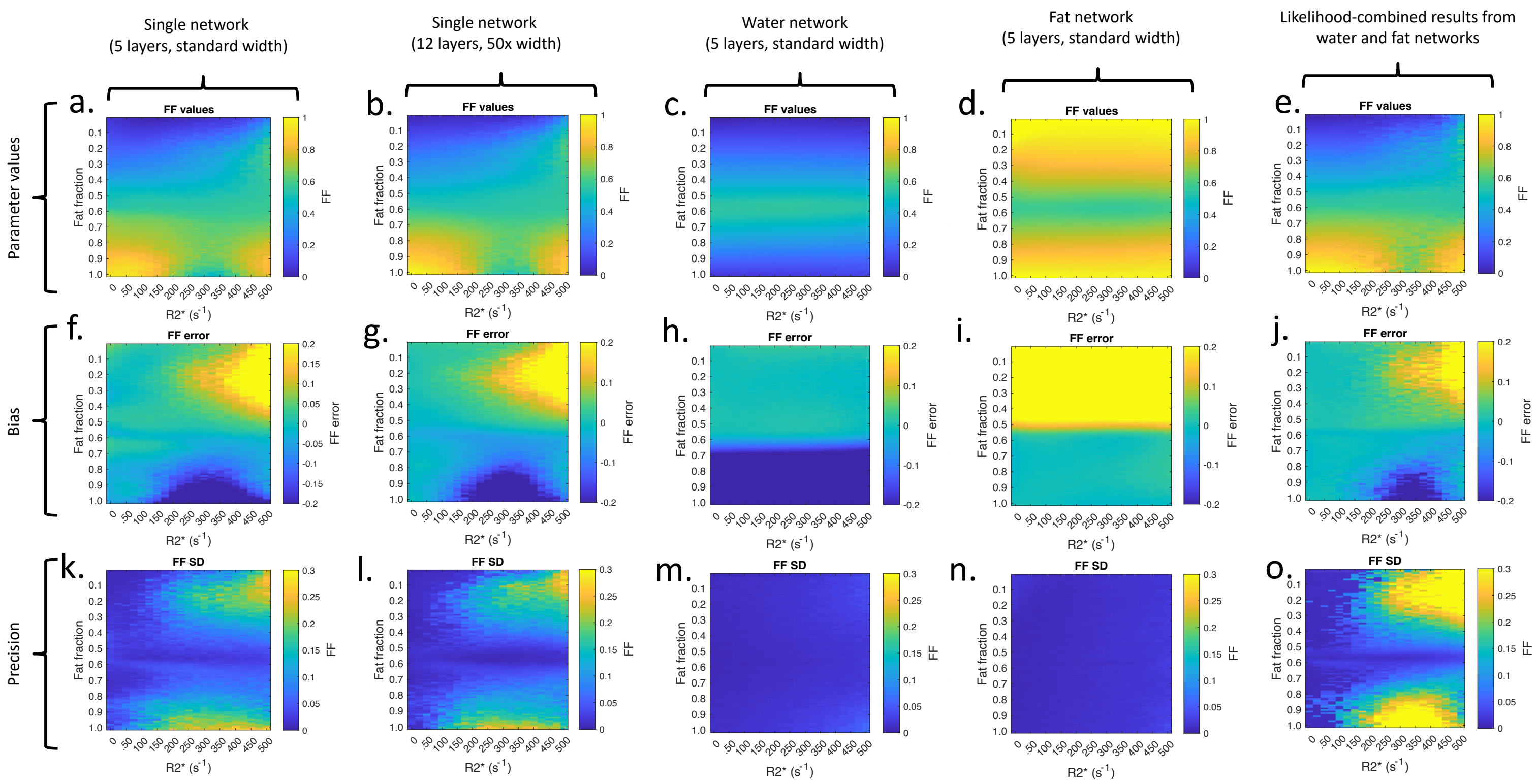

Figure 3: Improvement in PDFF estimation for dual network approach compared to single network (evaluation in the presence of noise). The figure shows the performance of a single network with no restriction of training distribution ((left and second columns, using standard and larger networks respectively), individual water and fat networks (third and fourth columns) and likelihood-combined networks (right column). The tissue parameter values are shown on the top row, bias is displayed in the middle row and tissue parameter standard deviation is shown on the bottom row.

Similarly, Figure 4 (left column) shows the $R_2^*$ estimation performance of a single network trained with a uniform distribution of tissue parameter values over the full tissue parameter space, evaluated in the presence of noise. At high PDFF, there are regions of positive error (shown in yellow) towards the bottom of subplot (f), and some regions of negative error (shown in blue) at low PDFF, as well as a band of negative error at high $R_2^*$. Again, the second column shows minimal improvement in the results for a substantially larger network.

Figure 5 shows individual parameter estimates obtained for two particular ground truth parameter combinations for the two different single network implementations.

For the full simulation experiment over 212100 signal instantiations, inference took 3.9s compared to 1649s (i.e. 27 minutes) for conventional fitting (423 fold acceleration).

### *4.1.2 Dual networks with modified training distribution*

*Water network*

The third column of Figure 3 shows the PDFF estimation performance of a single network trained with a restricted, low PDFF distribution of tissue parameter values (the 'water network'), evaluated in the presence of noise. Compared to the single network, there is a substantial improvement in performance for low PDFF values, with elimination of the high error, high variance regions at high $R_2^*$. However, as expected, the network fails on the high PDFF values as it has not experienced this region of tissue parameter space during training. Furthermore, (c) shows that there is a direct relationship between the bias in this region and the ground truth value, suggesting that this failure is a consequence of degeneracy.

Similarly, the third column of Figure 4 shows the $R_2^*$ estimation performance of the same water network. Compared to the single network, there is an improvement in performance for low PDFF values. However, as expected, the network fails on the high PDFF values as it has not experienced this region of tissue parameter space during training.

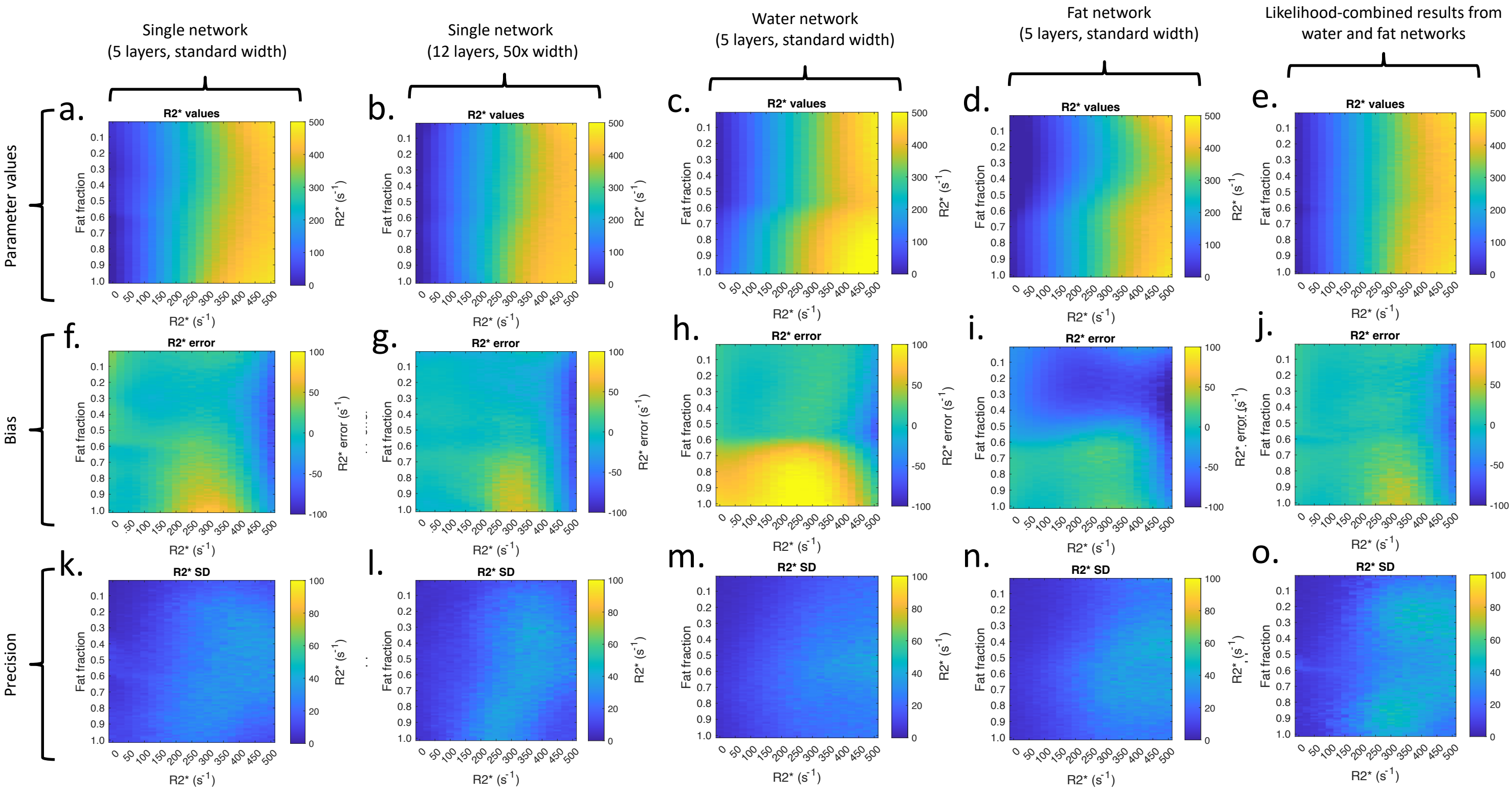

Figure 4: Improvement in $R_2^*$ estimation for dual network approach compared to single network (evaluation in the presence of noise). The figure shows the performance of a single network with no restriction of training distribution (left and second columns, using standard and larger networks respectively), individual water and fat networks (third and fourth columns) and likelihood-combined networks (right column). The tissue parameter values are shown on the top row, bias is displayed in the middle row and tissue parameter standard deviation is shown on the bottom row.

*Fat network*

The fourth column of Figure 3 shows the performance of a single network trained with a restricted, high PDFF distribution of tissue parameter values (the 'fat network'). Compared to the single network, there is a substantial improvement in performance for high PDFF values, particularly at high $R_2^*$. Again, as expected the network fails on the low PDFF values, and (d) shows a direct relationship between the bias in this region of tissue parameter space and the ground truth value, suggesting degeneracy as the underlying cause.

Similarly, the fourth column of Figure 4 shows the $R_2^*$ estimation performance of the same fat network. Compared to the single network, there is a substantial improvement in performance for high PDFF values. Again, as expected, the network fails on the low PDFF values as it has not been exposed to these values during training.

*Combined network outputs*

The right-hand column of Figure 3 shows the 'combined' performance of the two networks (i.e. the performance of the chosen output of the two networks based on likelihood). Compared to the single network (left column), the performance is substantially improved, with a substantial reduction in bias shown in (j). Importantly, the method retains its performance at substantially higher $R_2^*$ values than the single network method.

Note that the increase in tissue parameter standard deviation at high $R_2^*$ shown in the precision plot (o) compared to (k,l) reflects the fact that the tissue parameter estimates in (o) are obtained from a choice of two networks which are each accurate in their own region of tissue parameter space, whereas those in (k,l) are consistently biased towards the middle of the PDFF range across the tissue parameter space. This behaviour is shown explicitly in Figure 5. In other words, the pattern in (o) does not indicate poorer performance but rather the avoidance of consistently biased and less informative estimates.

Similarly, the right-hand column of Figure 4 shows the 'combined' performance of the two networks in terms of $R_2^*$ estimation performance. Compared to the single network (left column and second column), the performance is slightly improved, with a reduction in bias. Again, this behaviour is shown explicitly in Figure 5.

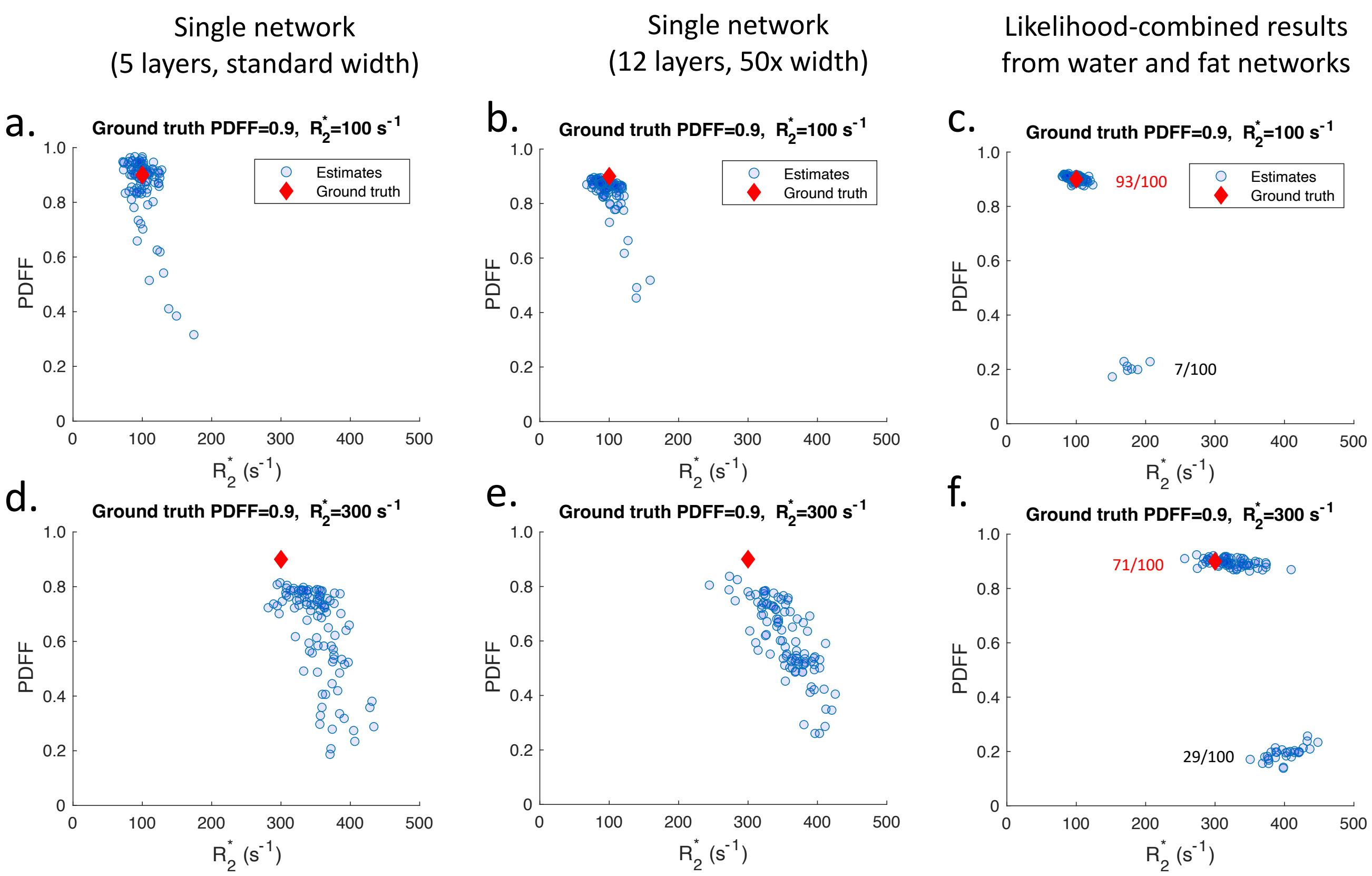

Figure 5: Scatterplot demonstrating differences in error pattern for single network methods compared to the dual-network RAIDER approach. Estimated parameters (blue markers) are compared to ground truth parameter values (red diamonds) for 100 noise instantiations. Results are shown for a small single network (left column), a much larger single network (middle column) and for the RAIDER approach (right column), for two different sets of ground truth parameters (top row and bottom row). Subplots (c) and (f) have also been annotated to show the number of estimates in the 'correct' and 'swapped' clusters (red and black text respectively). As expected from theory (Section 2.4), the single network estimates show 'averaging' behaviour and are scattered between the two optima, whereas the dual network estimates converge on one of the two likelihood optima.

*Comparison with conventional fitting*

Figure 6 shows the combined performance of the two networks in terms of PDFF bias and precision, compared against conventional fitting in the presence of noise. The performance of RAIDER is superior to conventional fitting with MAGORINO.

This figure also shows the combined performance of the two networks in terms of $R_2^*$ bias and precision. Again, the performance of RAIDER is similar to that of Rician magnitude fitting (with slightly higher bias, but also substantially improved precision, at high $R_2^*$).

For the full simulation experiment over 212100 signal instantiations, RAIDER inference took 3.7s, again compared to 1649s (i.e. 27 minutes) for conventional fitting (446 fold acceleration).

## 4.2 Phantom experiments

Results of the analysis of the multisite phantom data set are shown in Figure 7.

Median PDFF values for all 11 phantom vials are plotted against reference fat fraction values for all sites, acquisition protocols, and field strengths. There is good agreement between the predictions and the reference values with minimal bias across vendors and field strengths, and the overall performance was similar for the two methods. Some small differences were observed: in particular, for protocol 2 RAIDER performed better at some sites in the pure fat (high FF) vial.

Inference for the entire set of 28 phantom datasets (one slice for each) took 47.5s (i.e. 1.7s per dataset) for RAIDER, compared to 5.1 hours (i.e. 11 minutes per dataset) for MAGORINO (385-fold acceleration).

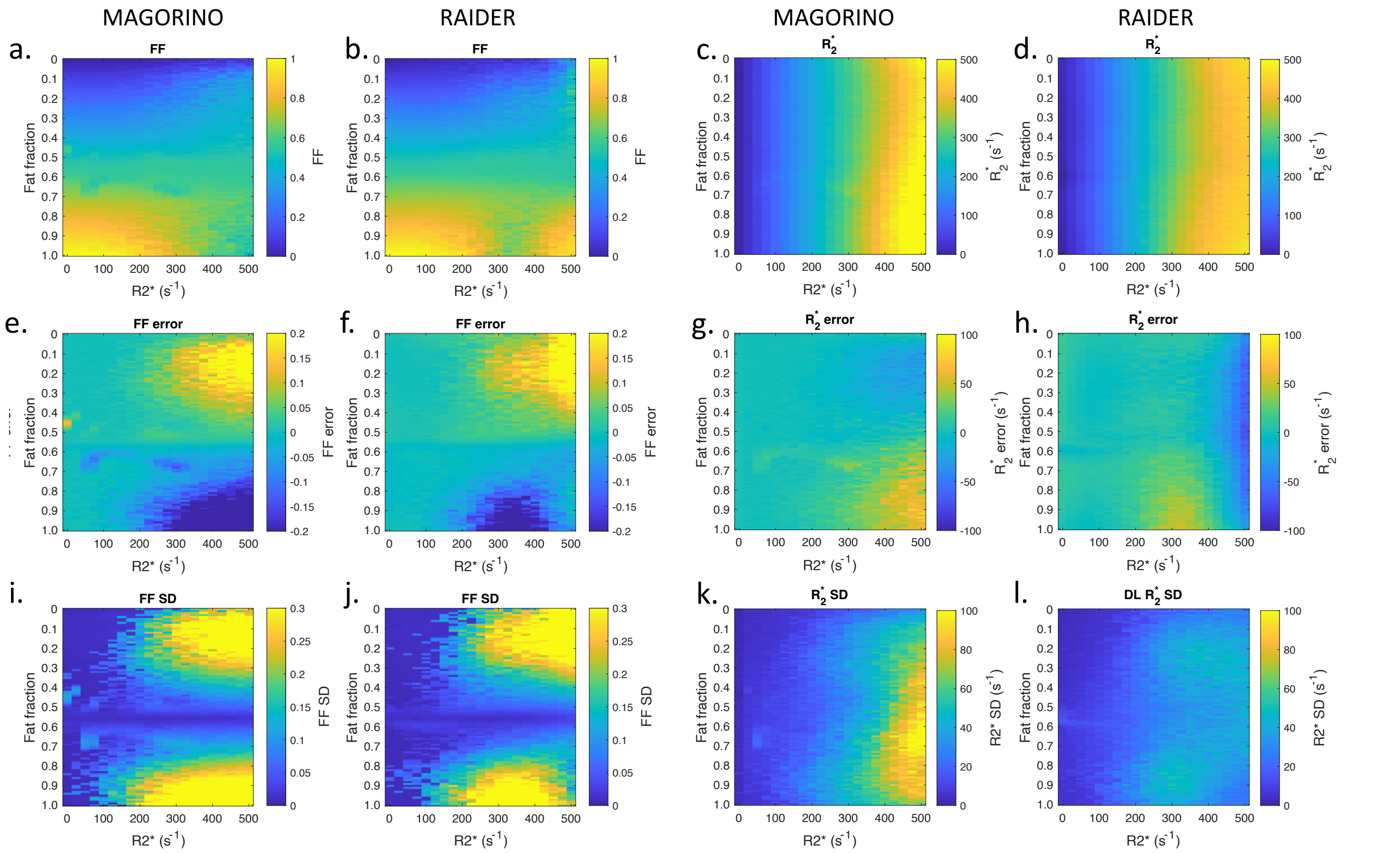

Figure 6: Performance of chosen output from two networks in the presence of noise and comparison against conventional fitting for PDFF (left two columns) and $R_2^*$ (right two columns). The top row shows estimated tissue parameter values, the middle row shows tissue parameter bias, and the bottom row shows tissue parameter variance. The first and third columns refer to conventional fitting using MAGORINO, and the second and fourth columns refer to RAIDER.

### 4.3 In vivo imaging

Example *in vivo* images are shown in Figures 8, 9, 10 and 11. Figure 8 illustrates how the outputs from the water network (a,e) and the fat network (b,f) are combined to produce the composite predictions shown in the third column (c,g). Figures 9 and 10 include difference maps to demonstrate differences compared to conventional fitting as well as the single network approach across multiple tissues, and also scatterplots showing these differences. The RAIDER performance shown in Figures 9 and 10 was superior to that for single network predictions, where high PDFF values showed substantial negative bias similar to that observed in simulations. Figure 11 includes images of the pelvis, hips and proximal thighs in the same subject as in Figures 9 and 10, providing a further demonstration of the ability of the algorithm to generalise to different anatomical regions.

Inference for the image slices shown in Figures 8, 9 / 10 and 11 took 1.2s, 0.7s and 0.9s respectively, compared to 56 minutes, 17 minutes and 18 minutes (2800-fold, 1457-fold and 1200-fold acceleration).

## 5. Discussion

Recently, several groups have explored the use of magnitude-based fitting for estimating PDFF and $R_2^*$ from chemical shift-encoded MRI data, since this approach can be used when complex-based methods fail or when phase data are inaccessible or unreliable, such as in multi-centre studies, low resource settings, preclinical imaging or when analysing large existing datasets, including biobank data. Magnitude fitting is also commonly used as a final processing step with a variety of complex-based methods. However, such methods are currently limited by their computational cost, increasing processing time and financial cost. Here, we propose RAIDER, which uses deep learning to separate water and fat in a 'voxel-independent fashion', performing tissue parameter estimation separately in each voxel. The proposed method resolves a key problem with degeneracy in magnitude-based voxelwise deep learning, enabling us to unlock the potential speed advantage of single-voxel magnitude-only. Furthermore, by providing separate water and fat networks, the RAIDER networks also enables the use of DL-based magnitude fitting to refine predictions of other methods within a multistep process, as is commonly done with conventional magnitude fitting (Zhong et al., 2014; Hernando et al., 2017), since the appropriate network (fat or water) could be chosen depending on the output of the previous step.

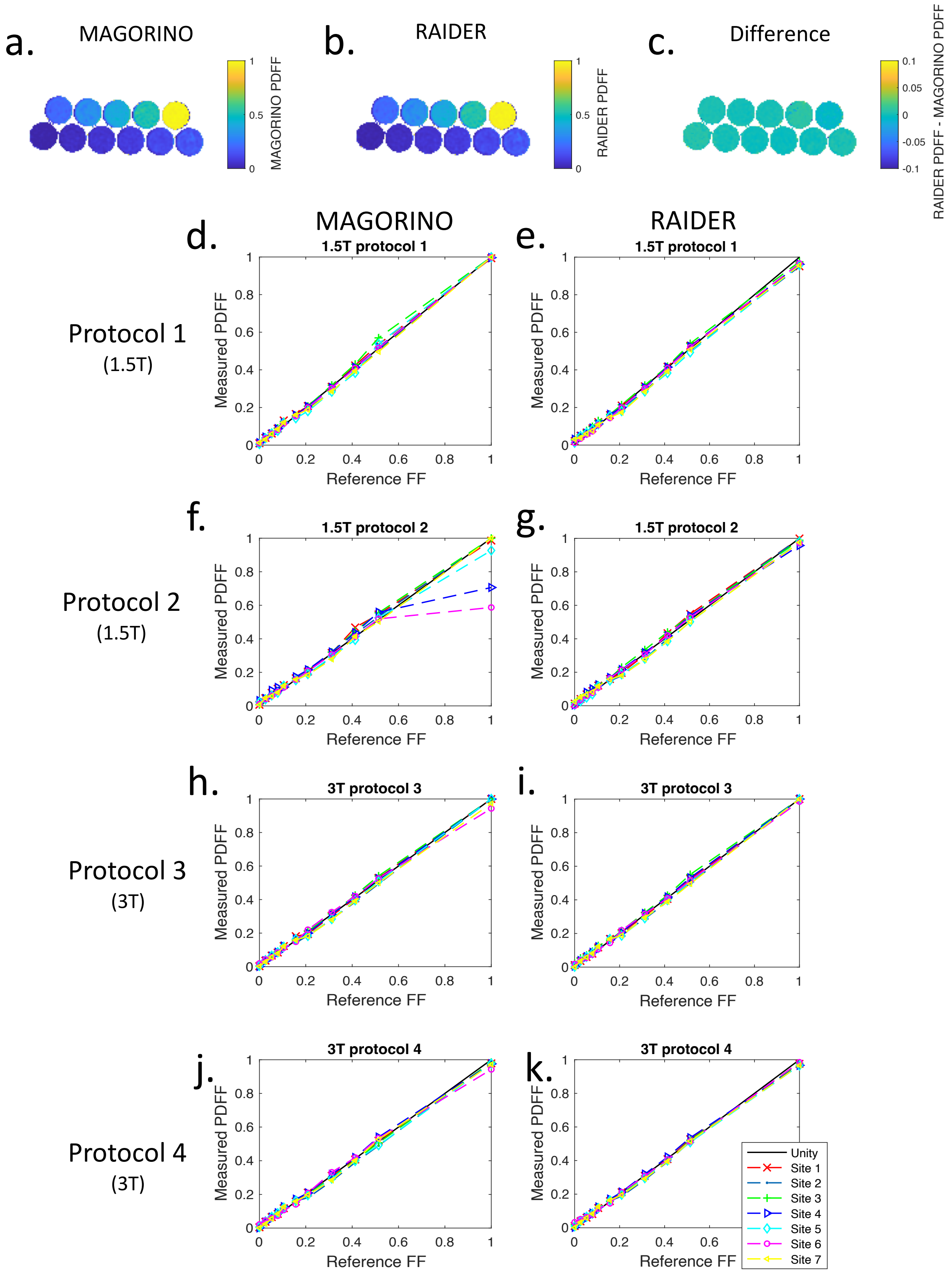

Figure 7: Results from multisite, multivendor, multi-field-strength phantom data set. Results from conventional fitting (MAGORINO) are shown in the left column, and RAIDER results are shown in the right column. Agreement plots are shown for each of the four protocols, with individual points for each of the six sites. The black 'unity' line indicates perfect agreement with reference PDFF values. "Site 7" refers to the repeat scans at site 1. The example image (top) is from site 1, 3T protocol 2, chosen to enable a direct visual comparison with corresponding illustrations in the MAGO and MAGORINO papers (Triay Bagur et al., 2018; Bray et al., 2023). Background voxels with very low signal intensity were set to white to facilitate interpretation. Reference FF values in the phantom were 0.026, 0.053, 0.079, 0.105, 0.157, 0.209, 0.312, 0.413, 0.514 and 1.

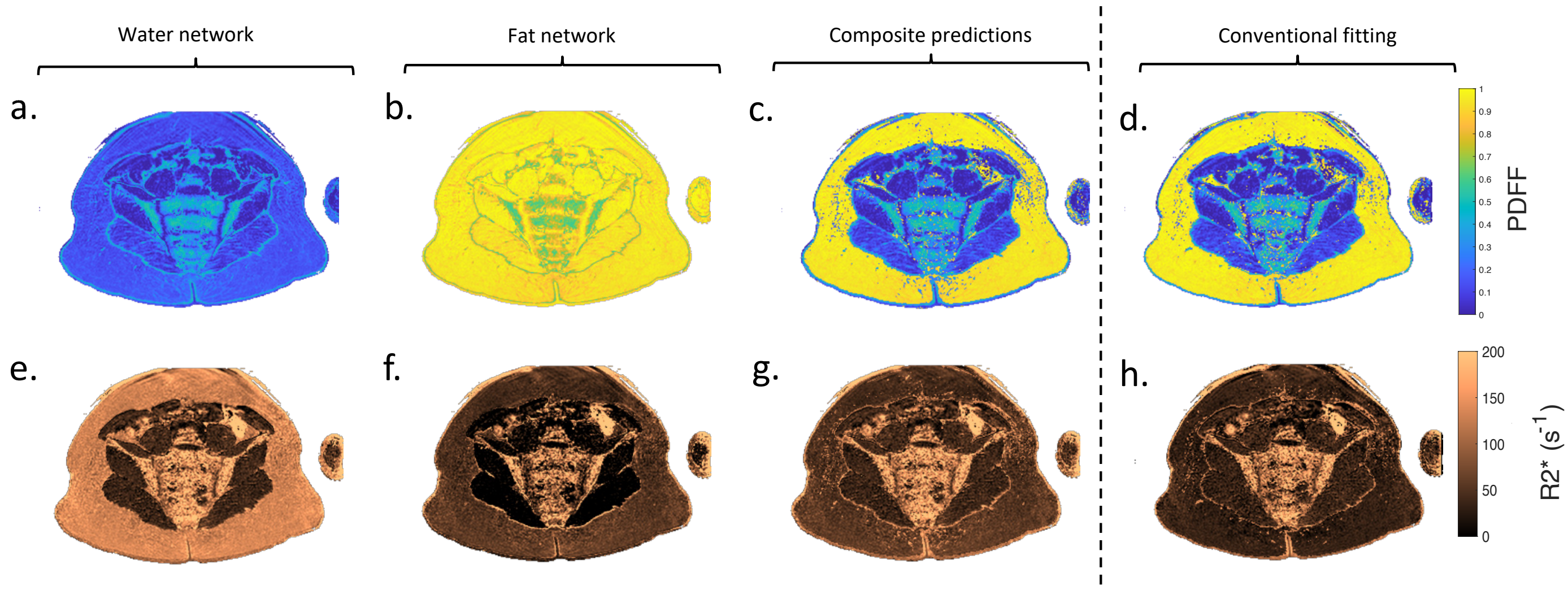

Figure 8: PDFF maps (top row) and $R_2^*$ maps (bottom row) from the pelvis of a volunteer. Note that the water network (first column) produces exclusively low PDFF estimates whereas the fat network (second column) produces exclusively high PDFF estimates. However, the chosen output from the two networks based on likelihood (third column) produces plausible predictions across a range of tissue types, including adipose tissue (almost pure fat), muscle (almost pure water) and bone marrow (approximately equal fat and water content). These agree closely with conventional fitting performed using MAGORINO (fourth column).

The proposed RAIDER approach, using two networks with separate training data distribution design, is motivated by our demonstration that using a single network with a full training distribution, including all plausible PDFF and $R_2^*$ values, produces relatively poor performance, with substantial bias in PDFF estimates. Furthermore, the observation that tissue parameter estimates are biased towards the mean (typically producing intermediate fat fraction values in voxels with high fat PDFF or low PDFF ground truths) is a strong indication that this problem is due to degeneracy (Guerreri et al., 2023). We showed that performance could be improved by training individual networks on a restricted distribution of tissue parameter values, specifically, by training one network only on low PDFF values (below the switching line) and another on high PDFF values (above the switching line). By selecting the most likely result from these networks, we were able to obtain predictions that outperformed a naive single network implementation across the full tissue parameter space, both in simulations and *in vivo*. Importantly, our results suggest a promising approach to resolving degeneracy in qMRI more broadly: the proposed method can be considered as analogous to methods using multiple start points in conventional fitting, and could be of value in any situation where this strategy is necessary to find the global optimum.

RAIDER offered similar or slightly improved performance to conventional MAGORINO fitting in terms of bias and precision, but was markedly faster, with an acceleration of 400-fold to 2800-fold. We also showed that this method could be successfully implemented in multisite, multivendor phantom data and *in vivo*. Our results suggest that the proposed method could help to reduce the time and expense associated with processing of CSE-MRI data for research and in clinical practice.

Thus far, for the purposes of fat-water separation, most authors have used CNNs, taking signals from the whole image as input, to separate fat and water (Goldfarb et al., 2019; Andersson et al., 2019; Jafari et al., 2021; Liu et al., 2021). These methods exploit the spatial relationships in the image as well as the information contained within individual voxels in the echo dimension, and typically use complex data. In addition to the need for complex data, these methods have other disadvantages: in particular, training is typically performed on real images from a specific anatomical region (with a particular resolution), meaning that large, expensive datasets are typically required to achieve satisfactory real-world performance. Furthermore, methods trained in this way are very unlikely to be fully generalizable to other regions of the body or to pathological cases that they have not been trained on. CNN-based methods are also highly vulnerable to differences in acquisition settings such as image matrix, and training networks using this approach is typically extremely slow, with training taking many hours or even days (Goldfarb et al., 2019; Andersson et al., 2019; Jafari et al., 2021; Liu et al., 2021). This limitation also applies to a related class of techniques referred to as model-based reconstruction methods, which aim to derive parameter maps from undersampled k-space data (thereby accelerating the acquisition), and have an additional requirement that the raw k-space data is available (Feng and Liu, 2020). In contrast, our RAIDER method has the advantage that there is no inherent dependence on anatomy, pathology or acquisition geometry introduced during training, making them intrinsically more generalizable across anatomical regions ('anatomy-independent'), as well as across clinical applications and scan geometries.

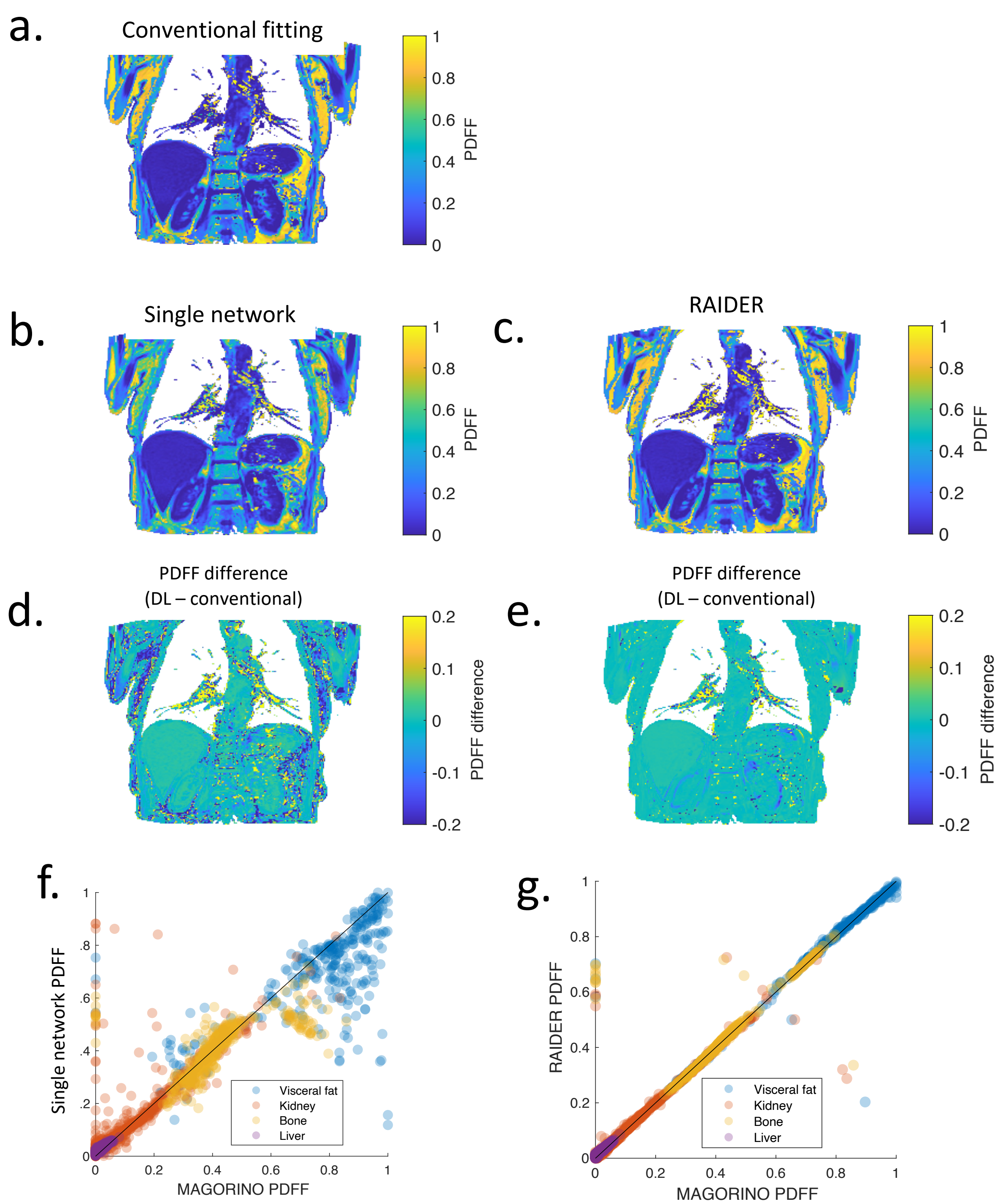

Figure 9: PDFF images of the thorax and abdomen. PDFF maps using conventional fitting (a), a single network DL implementation (b) and RAIDER (c) are shown. Difference maps between the single network approach and conventional fitting (d) and between RAIDER and conventional fitting (e), as well as scatterplots showing differences in key tissues (f,g) are also shown. For the single network method, there is substantial negative bias in the visceral fat and in the higher PDFF voxels in the bone marrow, matching the pattern expected from simulation; this error is substantially reduced for RAIDER. The precision of RAIDER estimates is also superior to that from the single network method. Note that very low signal voxels in the background were masked out and are displayed as white to increase the clarity of the images.

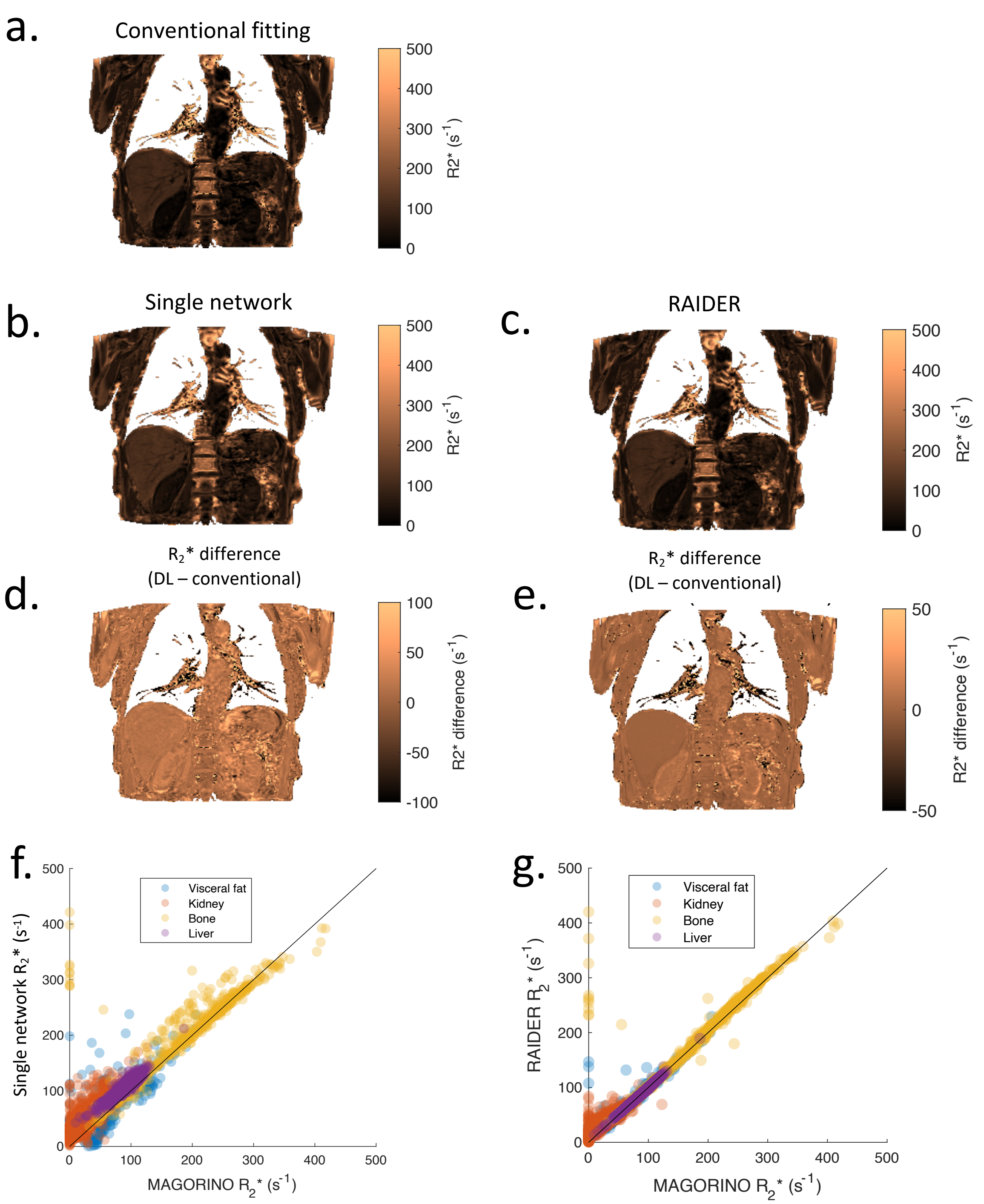

Figure 10: $R_2^*$ images of the thorax and abdomen in the same volunteer as in Figure 9. $R_2^*$ maps using conventional fitting (a), a single network DL implementation (b) and RAIDER (c) are shown. Difference maps between the single network approach and conventional fitting (d) and between RAIDER and conventional fitting (e), as well as scatterplots showing differences in key tissues (f,g) are also shown. For the single network method, there is greater positive bias at low $R_2^*$ than for RAIDER, matching the pattern expected from simulation. The precision of RAIDER estimates is also superior to that from the single network method; this is particularly pronounced for the bone marrow (an intermediate PDFF, high $R_2^*$ tissue), again matching the pattern seen in simulation. Note that very low signal voxels in the background were masked out and are displayed as white to increase the clarity of the images.

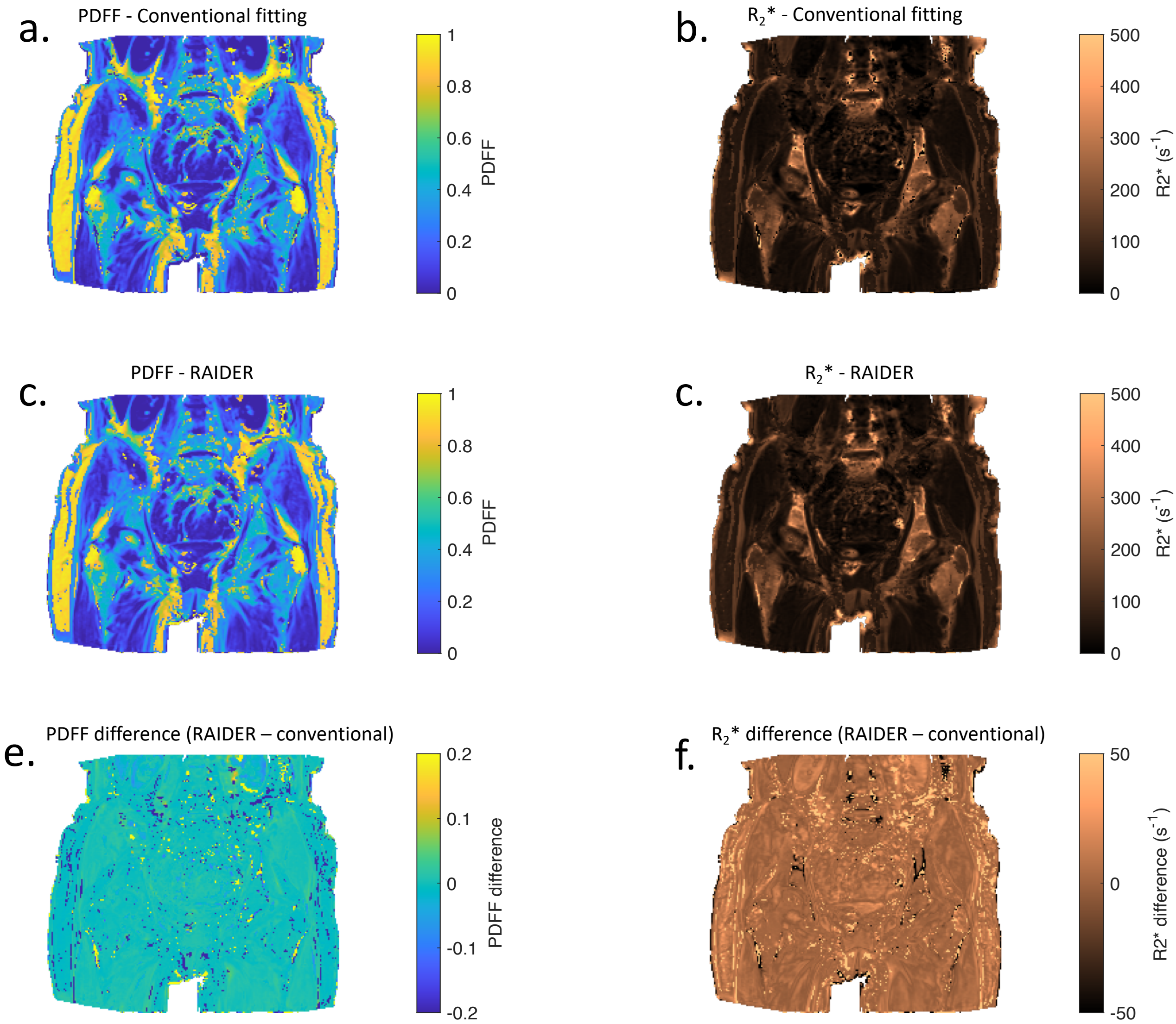

Figure 11: PDFF and$R_2^*$ maps of the pelvis, hips and proximal thighs in the same volunteer as in Figures 9 and 10, providing an additional demonstration of RAIDER to generalise to different anatomies. Conventional fitting results are shown on the top row, RAIDER results are shown in the middle row, and and difference maps are shown on the bottom row.

In this implementation, separate networks are required for different echo schemes, however, these networks can be trained relatively fast (using entirely simulated data, with no need for acquiring real data from many individuals) and are also compact, meaning that multiple versions of a model could very easily be stored on a scanner, having been trained in advance. This is a clear advantage over CNN-based methods, where different acquisition schemes will require new training data (which is orders of magnitude more expensive and difficult to obtain). Furthermore, RAIDER could potentially be extended by using methods such as neural controlled differential equations (Kuppens et al., 2024) to remove the need for re-training with different acquisition settings.

Further work is required to investigate the relative strengths and weaknesses of different deep learning-based methods, and to consider how hybrid approaches (for example, combining deep learning and conventional fitting methods) might be used to balance speed and performance. Overall, the potential acceleration offered by deep learning-based methods offers to substantially accelerate and simplify PDFF and $R_2^*$ estimation.

This work has some limitations. Firstly, under optimal conditions where the phase is completely reliable, the performance of magnitude-based methods, including the proposed RAIDER method as well as the existing MAGO and MAGORINO methods, is likely to be inferior to complex-based methods as well as more sensitive to model mismatch (Hernando et al., 2010; Wang et al., 2016). However, this class of methods is explicitly designed for situations where the phase is unavailable or unreliable, and for refining initial estimates produced by complex signal-based methods. As described in the Introduction, such magnitude-based approaches are used in a variety of settings and increasingly seen as a practical and generalizable method. Secondly, this is a 'proof-of-concept' study, describing a new methodological approach, and the evaluation with real data is preliminary; further work is required to evaluate real-world clinical utility and robustness (this would ideally be performed in multisite, multiscanner studies for each of the relevant clinical scenarios). Thirdly, in its present form the method may be vulnerable to domain shift, for example in tissues with low proton density or very high $R_2^*$ which are not encountered in the training data. This could be addressed by incorporating alternative DL approaches such as the self-supervised learning approach proposed by Barbieri et al. (2019). However, for most existing clinical applications this is of little concern, as the training method can easily be tailored to the tissues in question. Fourthly, the current implementation produced slightly biased $R_2^*$ estimates in subcutaneous fat, likely because the single peak normalization procedure (see 3.1.1) is imperfect and can result in more inaccurate $S_0$ estimates in this tissue due to fat-fat interference. The current method also shows a small negative bias at high $R_2^*$ values, again likely to be related to normalisation. These issues may be addressed in future work by further refinement of the normalisation procedure. Fifth, the choice of the 'switching point' is currently partly empirical (based on an understanding of the objective function shown in Figure 1), and there is scope for developing an analytical method for determining the precise value of this point. However, small changes in the switching point have very little little effect on performance. Finally, the final step of choosing between the outputs of the water and fat networks could potentially be improved by using another neural network to make the decision, although the current approach of likelihood comparison is well-founded in theory, and it is not clear whether a further neural network could improve on this without effectively introducing a prior (which might have its own drawbacks).

## 6. Conclusions

We propose RAIDER, a voxelwise method for **r**apid, **a**natomy-**i**ndependent **de**ep lea**r**ning-based $PDFF$ and $R_2^*$ estimation using multi-echo magnitude-data. RAIDER utilises two neural networks, each with a separately-designed training data distribution, to deal with degeneracy and thus realise the acceleration offered by an MLP-based approach, namely a 400-fold to 2800- fold acceleration. This means that processing takes just seconds, rather than minutes or hours.

## Acknowledgments

TJPB and MHC received personal support by the UCLH National Institute for Health Research Biomedical Research Centre (NIHR BRC); TJPB was also supported by NIHR BRC grants BRC1122/HEI/JP/110410 and BRC1185/III/TB/101350, and by an NIHR-Rosetrees Advanced Fellowship (code pending). This work was undertaken at UCLH/UCL, which receives funding from the UK Department of Health's NIHR BRC funding scheme. The funders had no role in study design, data collection and analysis, decision to publish, or preparation of the manuscript.

## Ethical Standards

The work follows appropriate ethical standards in conducting research and writing the manuscript, following all applicable laws and regulations regarding treatment of animals or human subjects.

## Conflicts of Interest

We declare we do not have any conflicts of interest.

## Data availability

The code for RAIDER implementation is available at `https://github.com/TJPBray/DixonDL`. The code used to generate the results in this paper are preserved as release v1.1 and are available in the Github repository, or under Zenodo DOI 10.5281/zenodo.17295346 ( `https://zenodo.org/records/17295346`).

## Appendix A. Noise-free results

Results for a noise-free version of the experiments in Section 3.2 are provided in Figure 12.

## Appendix B. Learning curves

Key training curves for the experiments performed in 3.2 and in Appendix A are provided in Figure 13.

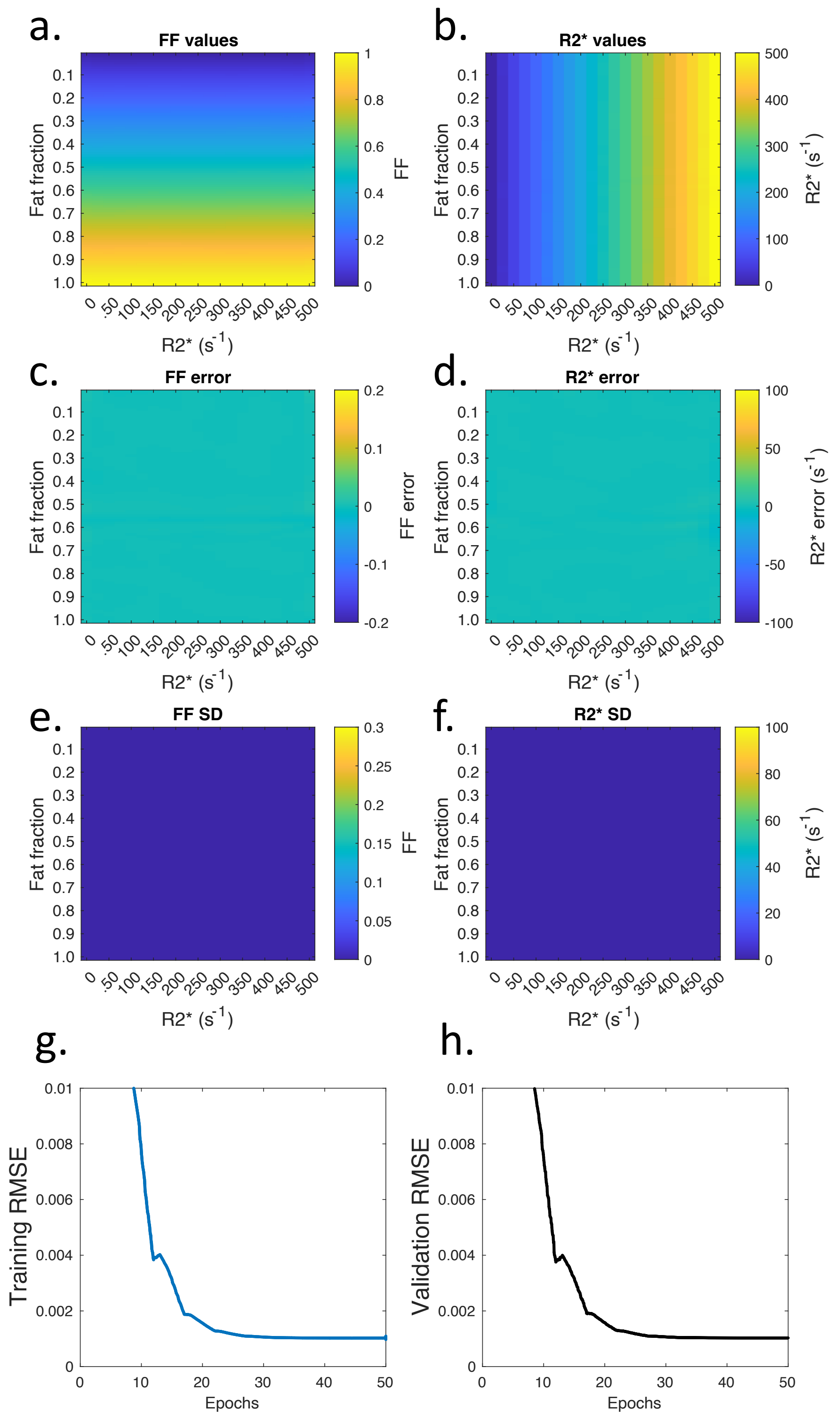

Figure 12: Performance of the larger single network (12 layers, 50x width) in with noise-free training and testing. There is very little error across the entire tissue parameter space, indicating that (i) there is minimal true degeneracy without noise and (ii) the network has sufficient capacity to describe the tissue model. The training curve for this network is displayed with very narrow y axis limits to show that there is no under or overfitting (this is the same curve as in the bottom left pane of Figure 11).

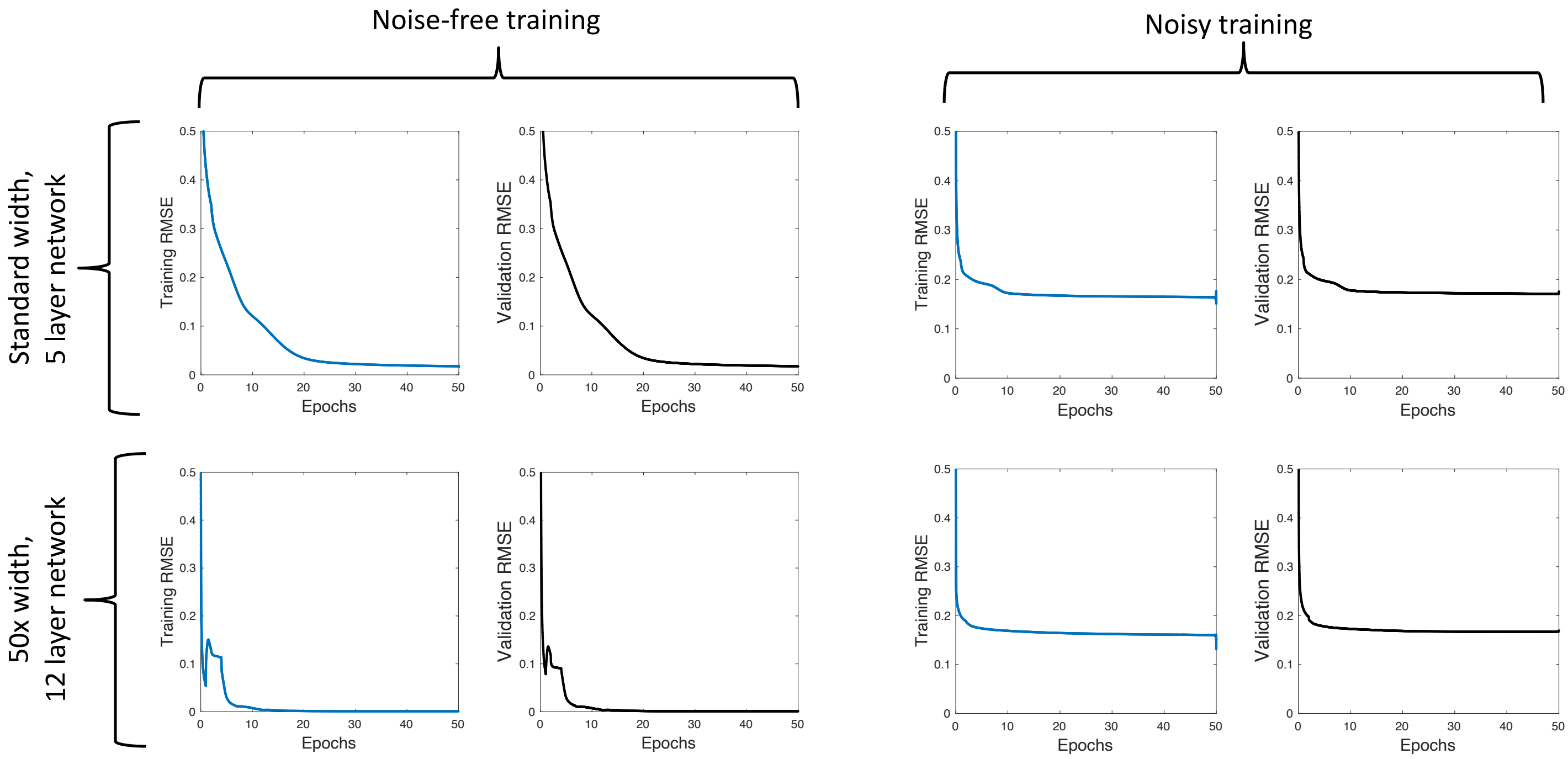

Figure 13: Training curves for standard width networks (top row) and larger networks (bottom row), with noise free data (left column) and with noisy data (right column). The training curve for the bottom left plots correspond to the performance plots shown in 12; the two plots in the right column correspond with the smaller and larger single networks whose performance is shown in 3 and 4 (first two columns of both figures). The performance plots for the top left training curves are not shown; this was slightly poorer than for the results in 12, in keeping with the higher RMSE on the training curves.